\documentclass[%
 reprint,
 superscriptaddress,
 amsmath,amssymb,
 aps,
]{revtex4-2}

\usepackage{graphicx}   
\usepackage{dcolumn}    
\usepackage{bm}         
\usepackage{verbatim}   
\usepackage{footnote}
\usepackage{subfigure}
\usepackage{xcolor}
\usepackage{float}
\usepackage[most]{tcolorbox}
\usepackage{soul}
\usepackage{natbib}
\usepackage[colorlinks=true,linkcolor=blue,allcolors=blue]{hyperref}
\usepackage[framemethod=TikZ]{mdframed}
\usepackage{tikz}       

\begin{document}

\title{Multiplexed Readout of Superconducting Qubits Using a 3D Re-entrant Cavity Filter}

\author{Mustafa Bakr}
 \email{mustafa.bakr@physics.ox.ac.uk} 
 \affiliation{Department of Physics, Clarendon Laboratory, University of Oxford, OX1 3PU, UK}
\author{Simone D. Fasciati}
 \affiliation{Department of Physics, Clarendon Laboratory, University of Oxford, OX1 3PU, UK}
\author{Shuxiang Cao}
 \affiliation{Department of Physics, Clarendon Laboratory, University of Oxford, OX1 3PU, UK}
 \author{Giulio Campanaro}
 \thanks{Present address: Alice and Bob. 53 Bd du Général Martial Valin 75015 Paris, France}
 \affiliation{Department of Physics, Clarendon Laboratory, University of Oxford, OX1 3PU, UK}
 \author{James Wills}
 \thanks{Present address: Oxford Quantum Circuits. Reading  RG2 9LH, UK}
 \affiliation{Department of Physics, Clarendon Laboratory, University of Oxford, OX1 3PU, UK}
 \author{Mohammed Alghadeer}
 \affiliation{Department of Physics, Clarendon Laboratory, University of Oxford, OX1 3PU, UK}
 \author{Michele Piscitelli}
\affiliation{Department of Physics, Clarendon Laboratory, University of Oxford, OX1 3PU, UK}
\author{Boris Shteynas}
 \thanks{Present address: Oxford Quantum Circuits. Reading RG2 9LH, UK}
 \affiliation{Department of Physics, Clarendon Laboratory, University of Oxford, OX1 3PU, UK}
 \author{Vivek Chidambaram}
 \thanks{Present address: UK National Quantum Computing Center. Didcot OX11 0QX, UK}
 \affiliation{Department of Physics, Clarendon Laboratory, University of Oxford, OX1 3PU, UK}
\author{Peter J. Leek}
 \email{peter.leek@physics.ox.ac.uk} 
 \affiliation{Department of Physics, Clarendon Laboratory, University of Oxford, OX1 3PU, UK}

\begin{abstract}
\vspace{0.3em} 
Hardware efficient methods for high fidelity quantum state measurements are crucial for superconducting qubit experiments, as qubit numbers grow and feedback and state reset begin to be employed for quantum error correction. We present a 3D re-entrant cavity filter designed for frequency-multiplexed readout of superconducting qubits. The cavity filter is situated out of the plane of the qubit circuit and capacitively couples to an array of on-chip readout resonators in a manner that can scale to large qubit arrays. The re-entrant cavity functions as a large-linewidth bandpass filter with intrinsic Purcell filtering. We demonstrate the concept with a four-qubit multiplexed device. 
\\
\end{abstract}

\maketitle


\section{Introduction}
In quantum computing, measurements are crucial for determining the outcome of computations~\cite{nielsen} and observing error syndromes in quantum error correction~\cite{ref1, ref2, ref3, ref4, ref5, ref6, ref7}. Superconducting circuits have successfully employed quantum non-demolition dispersive readout to measure qubit states~\cite{disread}. This technique uses the qubit-state-dependent frequency shift of a resonator, dispersively coupled to the qubit and probed with coherent microwave fields. Combining this technique with microwave parametric amplifiers~\cite{samp, peramp} has subsequently improved the speed and fidelity of the readout~\cite{50ns, 3Dmux, dis}. However, achieving both high readout speed and minimal qubit relaxation remains challenging due to competing requirements. Speeding up measurement requires increasing the resonator linewidth $\kappa$, which can adversely affect the qubit relaxation time $T_1$ due to enhanced spontaneous emission via the Purcell effect~\cite{pspeed}. Purcell filters are designed to mitigate this by engineering the electromagnetic environment surrounding the readout resonator to suppress qubit emission~\cite{fast1, fast2}, with recent implementations demonstrating readout fidelities above 99\% using a 40 ns readout pulse~\cite{40ns}. 

The frequency multiplexing of readout resonators can enable dispersive readout of multiple qubits using a single measurement cable~\cite{google, fid}. Various approaches to fast frequency-multiplexed readout in edge-connected circuits have been demonstrated, such as coupling multiple readout resonators to a single resonator Purcell filter~\cite{fast2}, or employing individual narrowband Purcell filters between the resonator and the output line~\cite{fid}. More recent off-device integration methods use distributed-element circuits~\cite{dis} or broadband stepped-impedance resonator filters on separate substrates~\cite{broadbandFF}. However, adding filters or circuit elements increases the device's complexity and footprint, which becomes a challenge when scaling to large numbers of qubits. Ongoing research is needed to integrate Purcell filters and multiplexing in a way that minimises decoherence channels, and avoids spurious modes, ensuring high-fidelity fast readout and long qubit relaxation times. The majority of available readout multiplexing designs rely on two-dimensional (2D) resonators and Purcell filters, e.g. in the form of coplanar waveguide (CPW) structures. An interesting approach recently demonstrated uses a CPW configuration to construct a readout resonator and filter resonator with auxiliary notch filtering, achieving 99.8\% fidelity in only 56~ns~\cite{Spring1}. Meanwhile, the use of fully three-dimensional (3D) electromagnetic modes~\cite{3Dmux} has remained largely unexplored, but could provide some useful advantages.

In this paper, we present a 3D re-entrant cavity designed for the multiplexed readout of superconducting qubits. This design capacitively couples to the qubit circuit out-of-plane, eliminating the need for direct galvanic connections and requires no additional space on the qubit chip itself. The modular nature of the system allows for straightforward adjustments to external coupling, enabling simple tuning of the resonator linewidth without requiring additional on-chip components. Furthermore, the cavity acts as a large line-width bandpass filter with a capacitive network that provides intrinsic Purcell protection. The coupling network between qubit, readout resonator, and filter can also be optimized to form compact notch filters, effectively eliminating Purcell decay through destructive interference across multiple paths. This intrinsic protection reduces the need for additional filtering components, simplifying the overall architecture and enhancing scalability for larger quantum systems. We demonstrate the concept in a four-qubit device, achieving average readout fidelity of 98.6\% and measurement-induced dephasing rates below 0.15 kHz within a measurement time of 1 $\mu$s, without the use of a parametric amplifier.

\begin{figure*}[!t]
    \centering
    \includegraphics[width=0.775\linewidth]{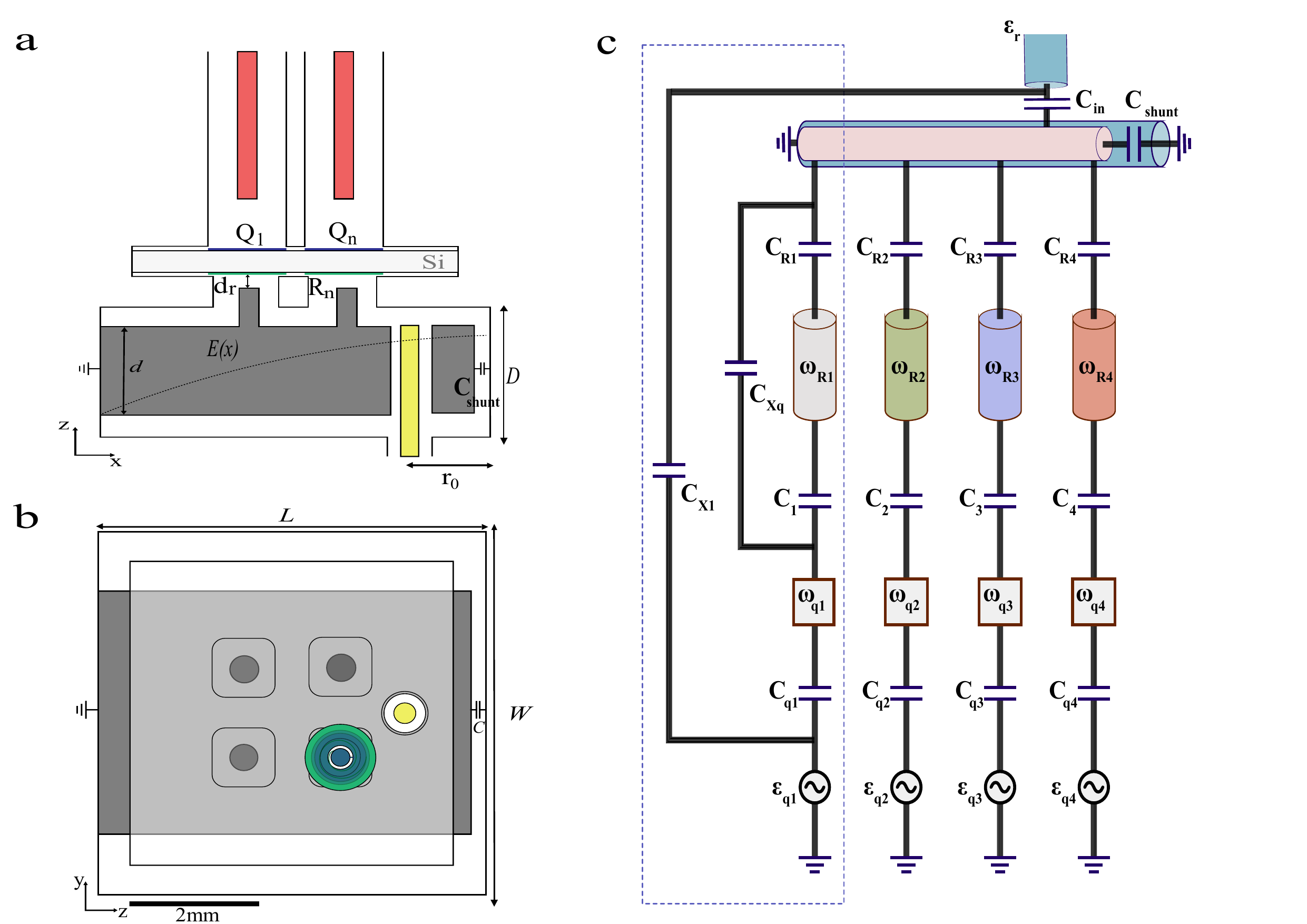}
    \caption{Side (a) and top (b) schematics of the device (false-colour cartoon), illustrating the 4:1 3D re-entrant cavity multiplexer. The device consists of a rectangular cavity with dimensions $L = 5.5 \, \text{mm}$, $d = 2 \, \text{mm}$, and $W = 4 \, \text{mm}$. The x-position of the common feedline coaxial cable from the enclosure ground, where the cavity is short-circuited, is $r_0 = 4.75 , \text{mm}$, and the distance between the extruded pins and readout resonators is $\rm d_{r} = 0.35 , \text{mm}$. In a, qubits (blue) are placed on the top side of the substrate (Si), readout resonators (green) on the bottom side, and the multiplexer is shown in grey. Qubits are addressed via capacitively coupled coaxial control lines (red), and readout resonators are addressed via the multiplexer common feedline (yellow). The electric field distribution $E(x)$ as a function of length is illustrated by the dotted line in a. Note that the qubit and resonator are shown only for one of the four qubits in (b) for clarity. (c) Conceptual circuit diagram for a multiplexed four-qubit device, showing an example interferometric circuit for qubit 1 ($\omega_{\rm q1}$) and readout resonator 1 ($\omega_{\rm r1}$). The circuit includes two cross-coupling paths through capacitors $C_{\rm x1}$ and $C_{\rm xq}$, introducing additional bandstop filtering notches at prescribed frequencies. The device comprises four qubits (angular frequencies $\omega_{\rm q1..q4}$) coupled to four readout resonators ($\omega_{\rm R1..R4}$) via coupling capacitors $C_{1..4}$. Each qubit is driven through its respective RF line and coupling capacitor $C_{\rm q1..q4}$. The filter resonator has an input capacitor $C_{\rm in}$ and a bandpass filter capacitor $C_{\rm shunt}$, and the entire system is characterized by an external quality factor $\epsilon_r$.
}
    \label{fig:mux}
\end{figure*}

\section{Design of a 3D Re-entrant Cavity}
Schematics of the out-of-plane readout multiplexer design are shown in Fig.~\ref{fig:mux}. The design is based around a rectangular cavity, within which is integrated a re-entrant rectangular section with four extruded pins extending vertically from one side. A coaxial port, with a characteristic impedance of $Z_{0} = 50$ ohms, capacitively couples to the re-entrant section of the cavity to route the readout signals via a UT47-type coaxial cable. The inner conductor of the coaxial cable extends into a through-hole in the re-entrant cavity. The re-entrant cavity is designed as a $\lambda/4$ filter with the electric field concentrated in the gap between the open end of the cavity and its metal enclosure. To reduce the physical size of the cavity to less than $\lambda/4$, the gap between the cavity's open end and the enclosure is reduced, modeled as a shunt-to-ground loading capacitor $C_{\rm shunt}$. The resonance is primarily determined by the cavity length $L$ and $C_{\rm shunt}$, while the internal quality factor is controlled by the ratio $D/d$. The array of extruded pins couples capacitively to the array of the readout resonators on the chip above, their capacitance being adjusted by varying the  gap $d_r$ between the pins and the resonators. For $r_{0} = 4.75$ mm and $\rm d_{r} = 0.35$ mm, the 3 dB bandwidth of the bandpass filter is 1.6 GHz at a center frequency of 9.8 GHz, with an external quality factor of 9. The designed linewidth of the readout resonators ($\kappa$) is 1.2 MHz. 

\begin{figure}[t!]
    \centering
    \includegraphics[width=\linewidth]{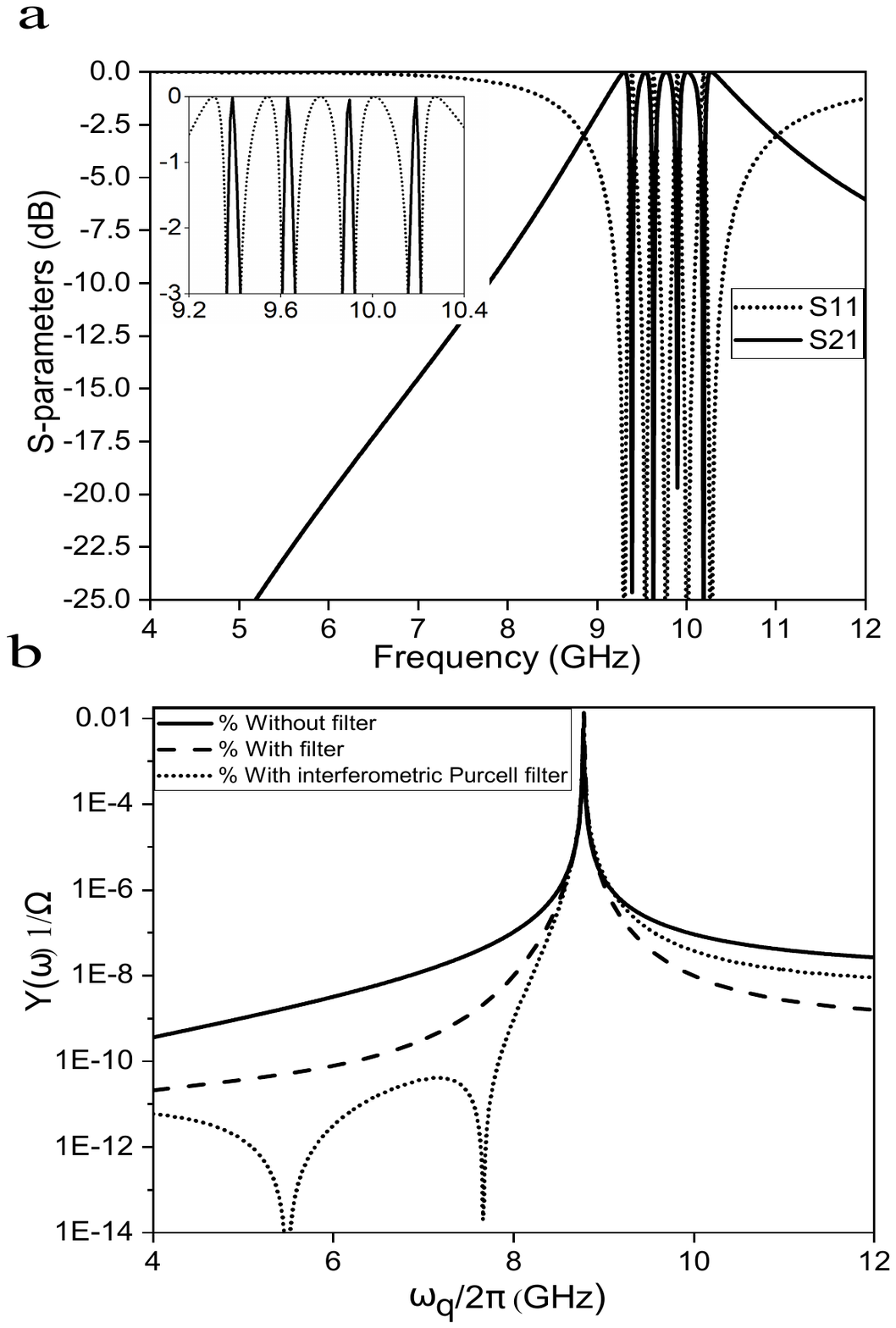}
    \caption{(a) Simulated scattering parameters for the 4:1 multiplexer, with geometry as shown in Fig.~\ref{fig:mux}. For the simulations, the single coupling port shown in Fig.~\ref{fig:mux}(b) is displaced by 1.5 mm along the y-axis and mirrored along the y-axis to perform a 2-port simulation and calculate the bandwidth of the filter. (Inset) Zoom of the scattering parameters centered at 9.8 GHz. (b) Simulated admittance of the environment seen from the Josephson junction (replaced by a high impedance port of one of the qubits) in the 4-qubit array under various configurations: without the multiplexer (solid line), with the multiplexer but no multiple-path interference (dashed line), and with both multiplexer and multi-path coupling described in the main text (dotted line).
}
    \label{fig:readout}
\end{figure}
In Fig.~\ref{fig:readout} we show the results of finite element simulations that demonstrate the behaviour of the above device. We first carry out a simulation of the scattering parameters for a modified design in which the single external port is replaced by a pair of ports (see Fig.~\ref{fig:readout}(a)). The multiplexer device acts as a broadband bandpass filter, providing Purcell filtering at the qubit frequency.  The design also enables the implementation of interferometric Purcell filters, as depicted in Fig.~\ref{fig:mux} (c), which feature non-adjacent capacitive couplings between the control coaxial inner connector, the re-entrant cavity and the qubit shunt capacitor, and the re-entrant cavity. The multiple coupling paths interfere destructively at the qubit frequency, creating bandstop filter notches (zeros of no transmission), which can enhance spontaneous emission suppression by approximately an order of magnitude. The frequency of these zeros is controlled by adjusting the coupling strength and signs of $C_{\rm x1}$ and $C_{\rm xq}$ in Fig.~\ref{fig:mux} (c). In Fig.~\ref{fig:readout} (b) we show the admittance of the environment seen by one of the qubits in the 4-qubit array with and without the cavity filter, and for the latter with and without the interferometric Purcell filter effect. 
\begin{figure}[!t]
    \centering
    \includegraphics[width=\linewidth]{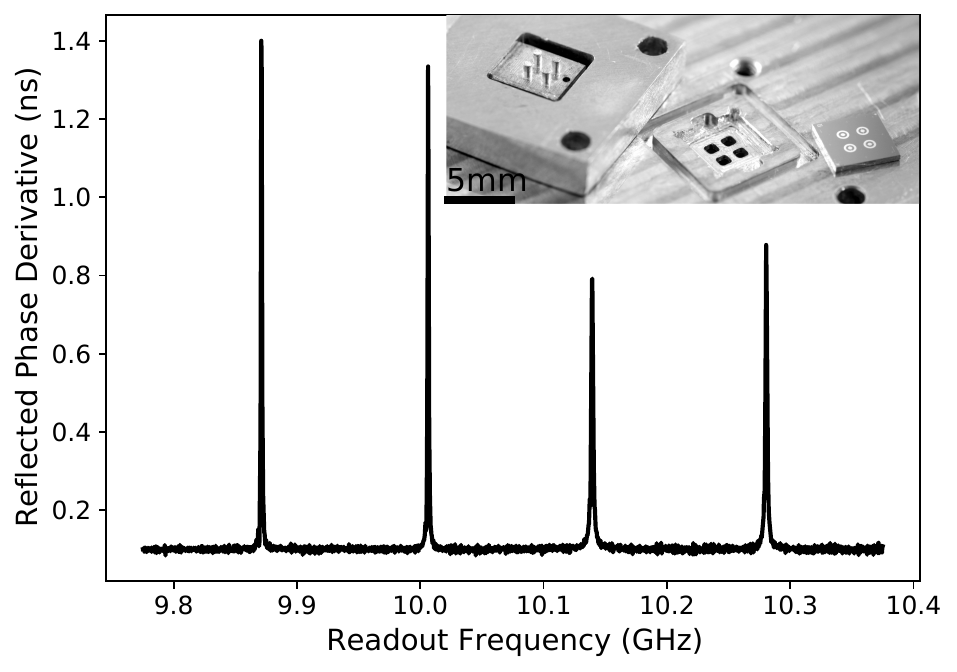}
    \caption{Reflection measurement of phase derivative (group delay ($\tau_D$)) at the feedline port of a 4:1 multiplexer coupled to a four-qubit device. The four readout resonances are clearly seen at 9.87, 10.01, 10.14, and 10.28 GHz. (Inset) Optical image of the micro-machined multiplexing cavity and a four-qubit device.} 
    \label{fig:reflection}
\end{figure}
In this unfiltered case, the readout resonator is directly coupled to the external port, without a bandpass filter or any multi-path interference effect. The introduction of a re-entrant cavity filter provides an order of magnitude suppression of Purcell decay compared to the unfiltered case. Further suppression is achieved by engineering a multipath cancellation capacitance network, with notch filters designed at 5.5 GHz and 7.7 GHz. In this configuration, the radius of the extruded pins of the re-entrant cavity filter, along with the geometry of the qubit and readout resonator capacitive electrodes, are optimized to adjust the strengths of $C_{\rm x1}$ and $C_{\rm xq}$, creating destructive interference at the prescribed frequencies. This effectively cancels out unwanted transmission paths that would otherwise contribute to Purcell decay. 

\section {Device Characterisation and Multiplexing Single Shot Readout}
We now describe measurements on a 4-qubit device incorporating the multiplexer described above. The device consists of four uncoupled flux-tunable coaxial transmons, with parameters outlined in Table \ref{tab:device1}. The device operates in a 3D-integrated coaxial architecture, with qubits and resonators fabricated on opposing sides of a silicon substrate \cite{Rahamim2017Double-sided, spring}. Characterisation of the device is performed with the flux-tunable transmons positioned at the flux sweet-spot. 
Fig.~\ref{fig:reflection} shows a measurement of the reflection spectrum at the feedline port, revealing four resonances close to the design frequencies of the readout resonators. By calculating the derivative of the phase with respect to frequency (group delay), we extract the resonator linewidths as 0.8, 0.9, 1.6, and 1.5 MHz through the relation \(\kappa_{\rm e}/2\pi = 2/(\pi \tau_{\rm D})\), where \(\tau_{\rm D}\) is the group delay. 

\begin{figure}[!t]
    \centering
    \includegraphics[width=\linewidth]{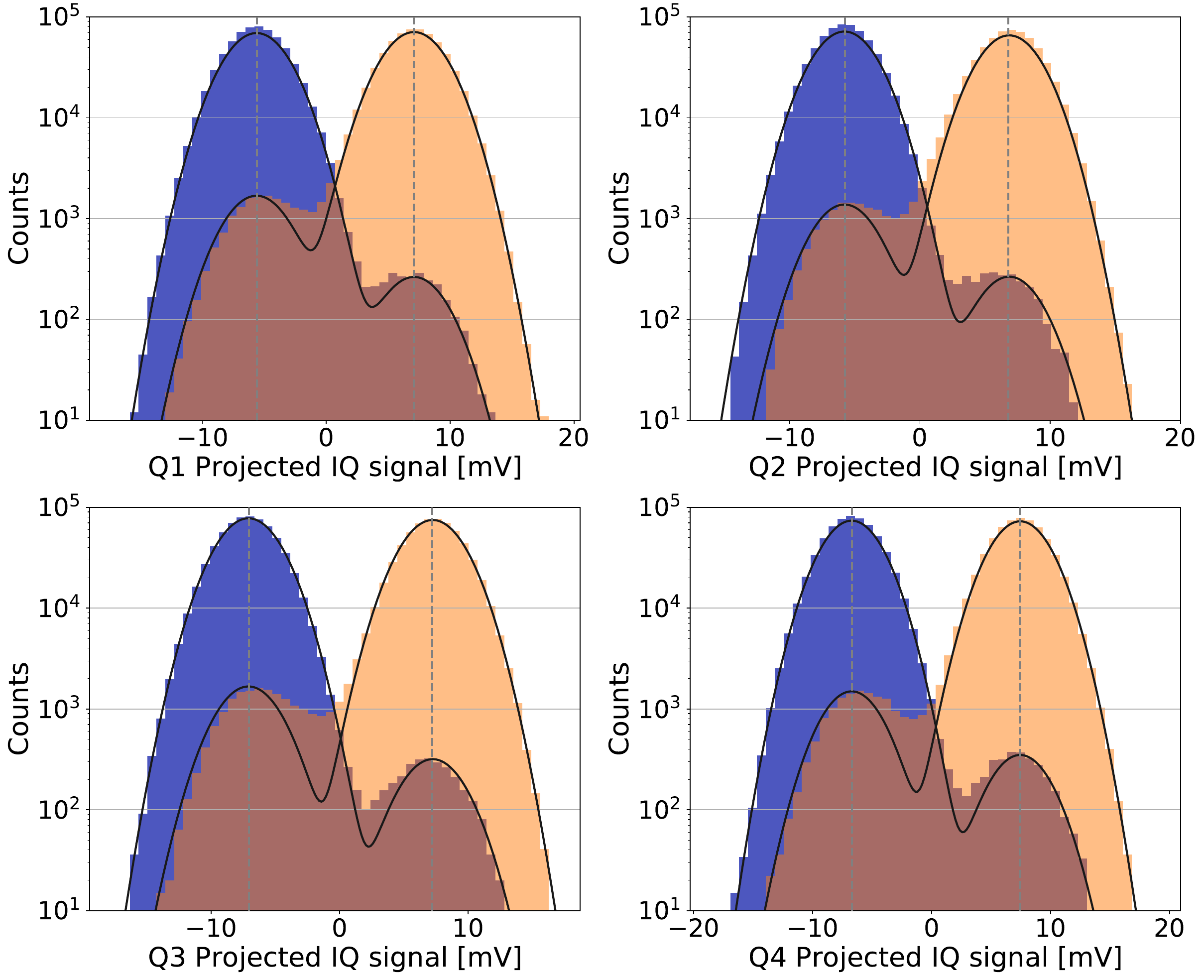}
    \caption{Histogram of integrated single-shot readout signals for Q1–Q4 with a $\pi$-pulse (orange) and without (blue), after heralding to remove thermal excitation in the initial state. Data fitted using a two-Gaussian model (solid lines) to represent the ground and excited state populations. The dotted lines show the projected means of the IQ signals for the ground and excited states after rotation into the 1D axis that maximizes signal contrast. Each state is prepared $\sim 6\times10^5$ times.}
    \label{fig:histogram}
\end{figure}

We next evaluate single-shot readout for each qubit by preparing the ground or excited state and applying readout pulses at the corresponding resonator frequencies. The readout integration time is set to 1 \(\mu\)s using a square pulse with 15 ns Gaussian edges for all resonators. We then optimise the readout amplitudes and frequencies individually to achieve the best signal-to-noise ratio (SNR) for qubits Q1 through Q4. We introduce a heralding measurement pulse before each cycle of qubit manipulation and measurement to remove thermal excitations in post-processing and ensure initialisation in the ground state~\cite{her}. The resulting complex signal from the qubit is downconverted to an intermediate frequency and digitised by an FPGA. This digitised signal is then projected onto a line bisecting the means of the two centroids in the IQ plane, corresponding to the qubit's ground and excited states. The final signal \(s\) for the two states, is displayed as a bimodal Gaussian distribution centered around $\bar{s}_0$ and $\bar{s}_\pi$ with characteristic width \(\sigma\), as depicted in Fig.~\ref{fig:histogram}. The signal-to-noise ratio (SNR) of the measurement is then calculated as \((\bar{s}_\pi - \bar{s}_0)/\sigma\), with values reported in Table \ref{tab:device2}.


\begin{table}[!t]
\centering
\begin{tabular*}{\columnwidth}{@{\extracolsep{\fill}} llllll}
\hline
Quantity & Unit & Q1 & Q2 & Q3 & Q4 \\
\hline
$\omega_{\rm q}/2\pi$ & GHz & 6.034 & 5.658 & 6.025 & 5.690 \\
$\alpha/2\pi$ & MHz & -221(1) & -218(1) & -221(1) & -217(1) \\
$\omega_{\rm R}/2\pi$ & GHz & 9.871 & 10.007 & 10.139 & 10.281 \\
$\kappa/2\pi$ & MHz & 0.8(1) & 0.9(1) & 1.6(1) & 1.5(1) \\
$\chi/2\pi$ & MHz & -1.7(1) & -1.3(1) & -1.5(1) & -1.3(1) \\
$T_1$ & $\mu$s & 49(8) & 50(7) & 52(7) & 47(7) \\
$T_2^*$ & $\mu$s & 23(3) & 37(5) & 33(4) & 39(6) \\
$T_{2E}$ & $\mu$s & 33(3) & 45(5) & 38(3) & 46(5) \\
$F_{\rm 1Q}$ & \% & 99.92 & 99.96 & 99.96 & 99.77 \\
\hline
\end{tabular*}
\caption{Parameters of the flux-tunable transmon device obtained from fits to reflection spectra and time-domain measurements: $\omega_{\rm q}/2\pi$ is the qubit frequency, $\alpha/2\pi$ is the anharmonicity, $\omega_{\rm R}/2\pi$ is the resonator frequency, $\kappa/2\pi$ is the resonator linewidth, $\chi/2\pi$ is the dispersive shift, $T_1$ is the relaxation time, $T_2^*$ is the Ramsey coherence time, $T_{2E}$ is the echo coherence time, and $F_{\rm 1Q}$ is the single-qubit (1Q) gate fidelity. Single qubit gate fidelities are measured using randomized benchmarking, separately on each qubit using the $\{ I, \pm X_{\frac{\pi}{2}}, \pm Y_{\frac{\pi}{2}}\}$ gate set. }
\label{tab:device1}
\end{table}

The bimodal Gaussian fits to the single-shot histograms highlight the sources of readout errors. The finite SNR leads to an overlap of the two Gaussian distributions, thereby impeding a complete distinction between the ground and excited states. With an average qubit energy relaxation time $T_1$ of approximately 50~$\mu$s and a signal integration time of 1~$\mu$s, we observe a readout error of approximately 1.5-2\% when preparing the excited state. This error is primarily caused by qubit decay during the measurement, which manifests as an increased ``shoulder'' in the histograms when the excited state is prepared. The thermal population of the qubits before heralding ranges between 9.5\% to 16\%, as reported in Table~\ref{tab:device2}. After post-selecting the data based on the heralding outcome, the residual excited state population is significantly reduced to between 0.2\% and 0.6\%. The remaining excitations are due to a combination of finite SNR and the possible re-excitation of a qubit during the 1~$\mu$s delay between the heralding pulse and the subsequent measurement sequence. This long delay is necessary for the ringdown of resonators with the smallest values of $\kappa$. We anticipate that the primary sources of error stem from $T_1$ decay during readout and thermal re-excitation post-heralding. The former could potentially be mitigated by adding a quantum-limited amplifier to shorten the measurement time, while the latter could be improved by increasing the $\kappa$ values to reduce the ringdown time, as well as through better filtering and thermalisation of the device to reduce excitations.

\begin{table}[!t]
\centering
\begin{tabular*}{\columnwidth}{@{\extracolsep{\fill}} llllll}
\hline
Symbol & Unit & Q1 & Q2 & Q3 & Q4 \\
\hline
SNR &  & 2.6 & 2.7 & 3.1 & 2.8 \\
$P_{\rm th}$ & \% & 10.1 & 9.5 & 10.9 & 15.7 \\
$P_{\rm post}$ & \% & 0.2 & 0.6 & 0.4 & 0.3 \\
$P(g|0)$ & \% & 99.2 & 99.4 & 99.5 & 99.4 \\
$P(e|\pi)$ & \% & 97.5 & 97.8 & 97.7 & 98.0 \\
$P_{\rm c}$ & \% & 98.4 & 98.6 & 98.6 & 98.7 \\
\hline
\end{tabular*}
\caption{State preparation and readout properties of the four-qubit device. 
SNR is the signal-to-noise ratio, $P_{\rm th}$ is the thermal population, and $P_{\rm post}$ is the post-heralding population. 
$P(g|0)$ and $P(e|\pi)$ denote the assignment probabilities of the ground state and excited state, respectively. 
$P_{\rm c}$ is the readout fidelity, measured during simultaneous single-shot readout of all four qubits.}
\label{tab:device2}
\end{table}

We classify the individual qubit states by assigning a binary value to each qubit measurement using a Gaussian Mixture Model (GMM) classifier, which is first applied to the combined 2D IQ plane data. The continuous-valued signal \(s\) from each qubit is then projected onto the optimal axis in the IQ plane, and the GMM fits two Gaussian distributions to distinguish between the ground and excited states. The readout fidelity of an individual qubit is quantified by the probability of a correct assignment, denoted as \(P_c = [P(g|0) + P(e|\pi)]/2\), where \(\pi\) indicates the state was prepared with a \(\pi\) pulse and 0 without a pulse. In this notation, \(e\) and \(g\) represent assignments to the excited and ground states, respectively. The obtained values of \(P(g|0)\), \(P(e|\pi)\), and \(P_c\) for each qubit are reported in Table \ref{tab:device2}. The average readout fidelity achieved is 98.6\%.

\begin{figure}[!t]
    \centering
    \includegraphics[width=\linewidth]{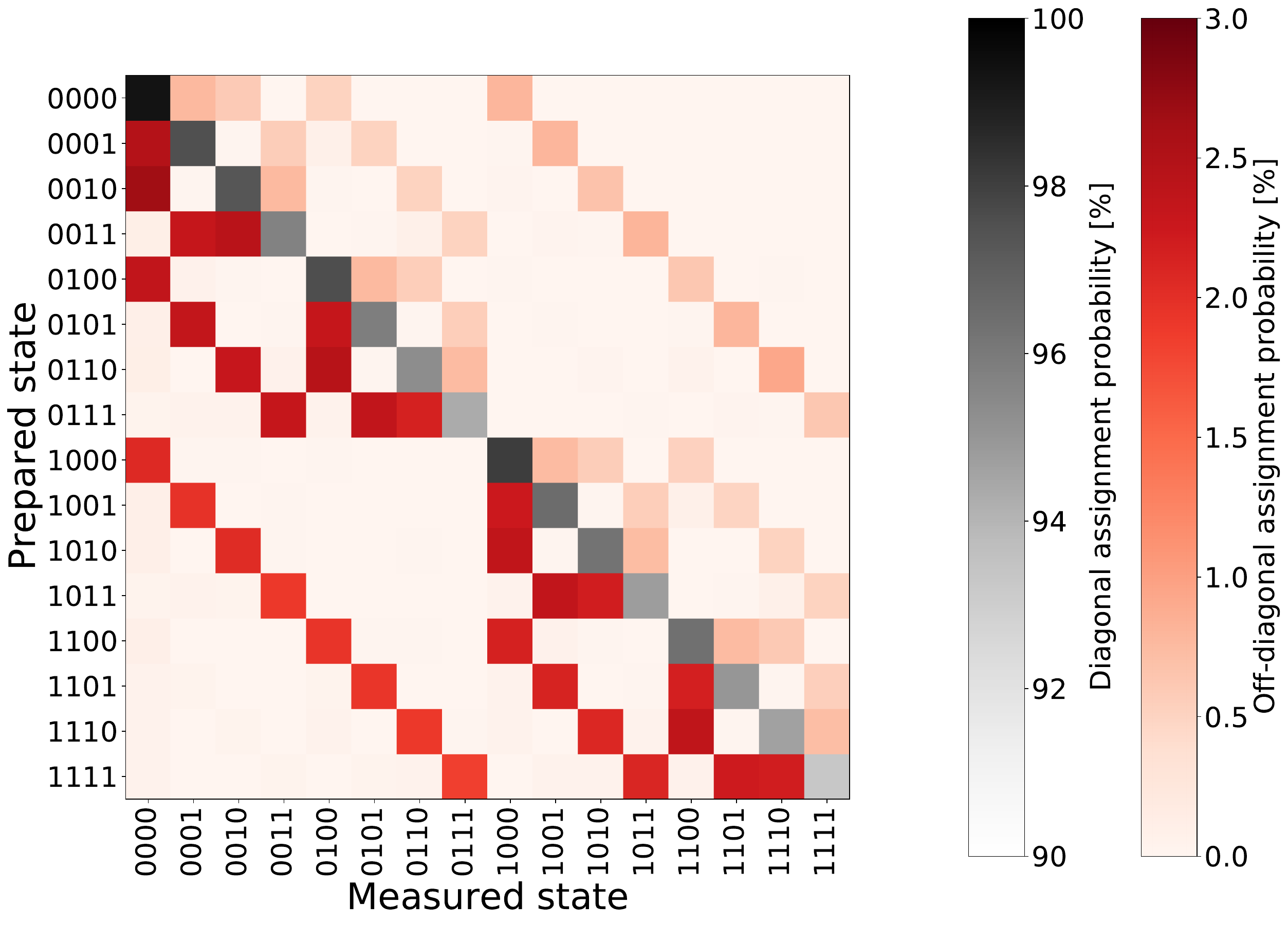}
    \caption{Assignment probability matrix for the four qubit device, with prepared state corresponding to each qubit initialised with a $\pi$-pulse or without a pulse. The qubits are ordered Q1, Q2, Q3, and Q4 from left to right (top to bottom) for the prepared (assigned) state. Each state is prepared $\sim10^6$ times.}
    \label{fig:ass}
\end{figure}

We also measure the assignment fidelity across all possible four-qubit states. Fig.~\ref{fig:ass} displays the resulting assignment probability matrix \(P(s_1 \ldots s_4|\zeta_1 \ldots \zeta_4)\) for all possible measurement outcomes with \(s_i \in \{g, e\}\) and prepared states with \(\zeta_i \in \{0, \pi\}\). The matrix derived from our experimental data shows that the average fidelity across all 16 states (diagonal elements) is 94.2\%. This is virtually identical to the prediction of 94.3\% obtained by multiplying the four $P_c$ values measured individually, showing that readout errors are largely uncorrelated.  Consistent with the previous observation that \(T_1\) decay is the main error source, we also find the highest and lowest fidelity for the global ground state, \(P(gggg|0000)\) = 97.3\%, and the global excited state, \(P(eeee|\pi\pi\pi\pi)\) = 91.3\%, respectively. The prominence of \(T_1\) decay error is visible in Fig.~\ref{fig:ass}, where the largest error contributions are all located below the diagonal (fewer excitations measured than prepared).  



\section{Characterization of Readout Crosstalk Via Measurement Induced Dephasing}
\begin{figure}[!t]
    \centering
    \includegraphics[width=\linewidth]{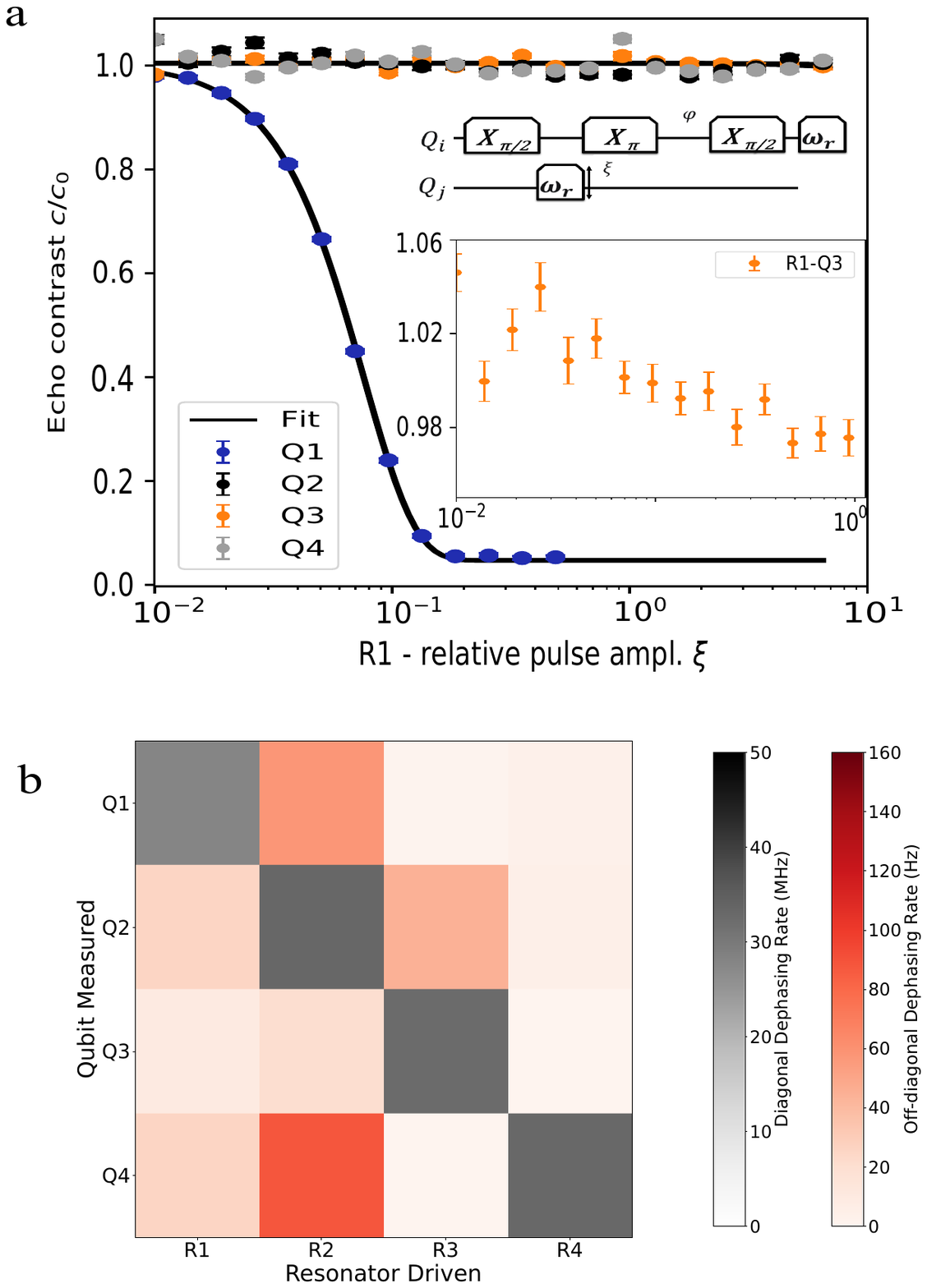}
    \caption{(a) Measurement-induced dephasing of qubits Q1–Q4 showing the relative echo oscillation contrast \( c/c_0 \) as a function of the readout pulse amplitude \( \xi \) on R1 normalised to the amplitude used for single-shot readout. The inset provides a zoomed-in view of the echo contrast for Q3 as a function of the relative readout pulse amplitude on R1. (b) Colour-map of the measurement-induced dephasing rates for each qubit-resonator pair across all readout resonators R1–R4. The fit errors for the dephasing rates are all bounded by 100 Hz (the highest being approximately 77 Hz). We approximately estimate bounds on measurement-induced dephasing rates to be less than 1 kHz.}
    \label{fig:echo}
\end{figure}

In a multiplexed readout setup, it is important not to inadvertently read out other qubits (readout crosstalk). In the case of our dispersive readout, this causes measurement-induced dephasing~\cite{dephasing}. In our architecture, there are two distinct potential causes of readout crosstalk. Firstly, this can occur via stray coupling between readout resonators and qubits in different unit cells of the circuit. Secondly, when driving readout resonators through the multiplexer, the finite spectral overlap of a readout pulse applied to resonator $R_i$ can drive a resonator $R_j$ if their frequencies are close enough. 
To quantify the impact of measurement crosstalk, we perform a Hahn-echo sequence experiment where a scaled readout pulse is inserted into the first half of the echo sequence, as depicted in the upper inset of Fig. \ref{fig:echo} (a). The interval between pulses is fixed, and the phase of the final $\pi/2$ pulse is swept to observe coherent oscillations. The decay of the echo oscillation contrast \(c(\xi)\) as a function of the relative readout pulse amplitude \(\xi\) is fitted to the model \(c(\xi) = c_0 \exp(-\Gamma \tau_p \xi^2)\), where \(c_0\) represents the initial contrast, \(\Gamma\) is the measurement-induced dephasing rate, and \(\tau_p\) is the total duration of the readout pulse~\cite{fid}. We extract the effective dephasing rates for each qubit-resonator pair to assess the crosstalk between untargeted qubits. Fig.~\ref{fig:echo} (a) and (b) illustrate the measurement-induced dephasing rates for both diagonal (self-dephasing) and off-diagonal (crosstalk) elements across the device. For qubit-resonator pairs with \(i \neq j\), the maximum measurement-induced dephasing observed is less than 0.15 kHz, which is negligible compared to qubit decoherence rates.

\section{SUMMARY}
In this paper, we introduce a 3D re-entrant cavity filter for multiplexed readout of superconducting qubits, designed to be out-of-plane and capacitively coupled to readout resonators. Through finite element simulations, we demonstrate the intrinsic Purcell filtering behavior of the multiplexer, mitigating qubit decay during readout. Our experimental results showcase a four-qubit device show an average readout fidelity of 98.6\% within a 1~$\mu$s integration time without the use of a parametric amplifier, with measurement-induced dephasing rates below 0.15~kHz. The compact 3D structure reduces the need for on-chip filtering components, simplifying integration.

A potential next step involves extending the multiplexer to accommodate larger multi-qubit systems. By incorporating multi-pole filtering sections into the re-entrant cavity structure, the design could capacitively couples to larger arrays of qubits while improving Purcell protection. This approach could scale to systems with tens or even hundreds of qubits while maintaining high readout fidelity

\begin{acknowledgments}
M.B. acknowledges support from EPSRC QT Fellowship grant EP/W027992/1. P. L. acknowledges support from [EP/T001062/1, EP/N015118/1, EP/M013243/1]. S.C. acknowledges support from Schmidt Science. 
\end{acknowledgments}



\begin{thebibliography}{99}

\bibitem{nielsen}
M. A. Nielsen and I. L. Chuang, \textit{Quantum Computation and Quantum Information}, Cambridge University Press, Cambridge, 2000.

\bibitem{ref1}
C. Ryan-Anderson, J. G. Bohnet, K. Lee, D. Gresh, A. Hankin, J. P. Gaebler, D. Francois, A. Chernoguzov, D. Lucchetti, N. C. Brown, T. M. Gatterman, S. K. Halit, K. Gilmore, J. A. Gerber, B. Neyenhuis, D. Hayes, and R. P. Stutz, Realization of real-time fault-tolerant quantum error correction, \textit{Phys. Rev. X} \textbf{11}, 041058 (2021).

\bibitem{ref2}
S. Krinner, N. Lacroix, A. Remm, A. Di Paolo, É. Genois, C. Leroux, C. Hellings, S. Lazăr, F. Swiadek, J. Herrmann, G. J. Norris, C. K. Andersen, M. Muller, A. Blais, C. Eichler, and A. Wallraff, Realizing repeated quantum error correction in a distance-three surface code, \textit{Nature} \textbf{605}, 669 (2021).

\bibitem{ref3}
R. Acharya \textit{et al.}, Suppressing quantum errors by scaling a surface code logical qubit, \textit{Nature} \textbf{614}, 676 (2022).

\bibitem{ref4}
V. V. Sivak, A. Eickbusch, B. Royer, S. Singh, I. Tsioutsios, S. Ganjam, A. Miano, B. L. Brock, A. Z. Ding, L. Frunzio, S. M. Girvin, R. J. Schoelkopf, and M. H. Devoret, Real-time quantum error correction beyond break-even, \textit{Nature} \textbf{616}, 50–55 (2023).

\bibitem{ref5}
D. Bluvstein, S. J. Evered, A. A. Geim, S. H. Li, H. Zhou, T. Manovitz, S. Ebadi, M. Cain, M. Kalinowski, D. Hangleiter, J. P. Bonilla Ataides, N. Maskara, I. Cong, X. Gao, P. S. Rodriguez, T. Karolyshyn, G. Semeghini, M. J. Gullans, M. Greiner, V. Vuletić, and M. D. Lukin, Logical quantum processor based on reconfigurable atom arrays, \textit{Nature} \textbf{626}, 58 (2023).

\bibitem{ref6}
A. Paetznick, M. P. da Silva, C. Ryan-Anderson, J. M. Bello-Rivas, J. P. Campora III, A. Chernoguzov, J. M. Dreiling, C. Foltz, F. Frachon, J. P. Gaebler, T. M. Gatterman, L. Grans-Samuelsson, D. Gresh, D. Hayes, N. Hewitt, C. Holliman, C. V. Horst, J. Johansen, D. Lucchetti, Y. Matsuoka, M. Mills, S. A. Moses, B. Neyenhuis, A. Paz, J. Pino, P. Siegfried, A. Sundaram, D. Tom, S. J. Wernli, M. Zanner, R. P. Stutz, and K. M. Svore, Demonstration of logical qubits and repeated error correction with better-than-physical error rates, arXiv preprint, arXiv:2404.02280 (2024).

\bibitem{ref7}
R. Acharya, L. Aghababaie-Beni, I. Aleiner, T. I. Andersen, M. Ansmann, F. Arute, K. Arya, A. Asfaw, N. Astrakhantsev, J. Atalaya, R. Babbush, D. Bacon, B. Ballard, J. C. Bardin, J. Bausch, A. Bengtsson, A. Bilmes, S. Blackwell, S. Boixo, \textit{et al.}, Quantum error correction below the surface code threshold, arXiv preprint, arXiv:2408.13687v1 (2024).


\bibitem{disread}
A. Wallraff, D. I. Schuster, A. Blais, L. Frunzio, J. Majer, M. H. Devoret, S. M. Girvin, and R. J. Schoelkopf, Approaching unit visibility for control of a superconducting qubit with dispersive readout, \textit{Phys. Rev. Lett.} \textbf{95}, 060501 (2005).

\bibitem{samp}
J. Aumentado, Superconducting parametric amplifiers: The state of the art in Josephson parametric amplifiers, \textit{IEEE Microwave Magazine} \textbf{21}, 45 (2020).

\bibitem{peramp}
M. Esposito, A. Ranadive, L. Planat, and N. Roch, Perspective on traveling wave microwave parametric amplifiers, \textit{Appl. Phys. Lett.} \textbf{119}, 120501 (2021).

\bibitem{50ns}
T. Walter, P. Kurpiers, S. Gasparinetti, P. Magnard, A. Potočnik, Y. Salathé, M. Pechal, M. Mondal, M. Oppliger, C. Eichler, and A. Wallraff, Rapid high-fidelity single-shot dispersive readout of superconducting qubits, \textit{Phys. Rev. Appl.} \textbf{7}, 054020 (2017).

\bibitem{3Dmux}
S. Kundu, N. Gheeraert, S. Hazra, T. Roy, K. V. Salunkhe, M. P. Patankar, and R. Vijay, Multiplexed readout of four qubits in 3D circuit QED architecture using a broadband Josephson parametric amplifier, \textit{Appl. Phys. Lett.} \textbf{114}, 172601 (2019).

\bibitem{dis}
Y. Sunada, S. Kono, J. Ilves, S. Tamate, T. Sugiyama, Y. Tabuchi, and Y. Nakamura, Fast readout and reset of a superconducting qubit coupled to a resonator with an intrinsic Purcell filter, \textit{Phys. Rev. Appl.} \textbf{17}, 044016 (2022).

\bibitem{pspeed}
A. A. Houck \textit{et al.}, Controlling the spontaneous emission of a superconducting transmon qubit, \textit{Phys. Rev. Lett.} \textbf{101}, 080502 (2008).

\bibitem{fast1}
M. D. Reed, B. R. Johnson, A. A. Houck, L. DiCarlo, J. M. Chow, D. I. Schuster, L. Frunzio, and R. J. Schoelkopf, Fast reset and suppressing spontaneous emission of a superconducting qubit, \textit{Appl. Phys. Lett.} \textbf{96}, 203110 (2010).

\bibitem{fast2}
E. Jeffrey, D. Sank, J. Y. Mutus, T. C. White, J. Kelly, R. Barends, Y. Chen, Z. Chen, B. Chiaro, A. Dunsworth, A. Megrant, P. J. J. O'Malley, C. Neill, P. Roushan, A. Vainsencher, J. Wenner, A. N. Cleland, and J. M. Martinis, Fast accurate state measurement with superconducting qubits, \textit{Phys. Rev. Lett.} \textbf{112}, 190504 (2014).

\bibitem{40ns}
Y. Sunada, K. Yuki, Z. Wang, T. Miyamura, J. Ilves, K. Matsuura, P. A. Spring, S. Tamate, S. Kono, and Y. Nakamura, Photon-noise-tolerant dispersive readout of a superconducting qubit using a nonlinear Purcell filter, \textit{PRX Quantum} \textbf{5}, 010307 (2024).

\bibitem{google}
Y. Chen, D. Sank, P. O'Malley, T. White, R. Barends, B. Chiaro, J. Kelly, E. Lucero, M. Mariantoni, A. Megrant, C. Neill, A. Vainsencher, J. Wenner, Y. Yin, A. N. Cleland, and J. M. Martinis, Multiplexed dispersive readout of superconducting phase qubits, \textit{Appl. Phys. Lett.} \textbf{101}, 182601 (2012).

\bibitem{fid}
J. Heinsoo, C. K. Andersen, A. Remm, S. Krinner, T. Walter, Y. Salathé, S. Gasparinetti, J.-C. Besse, A. Potočnik, A. Wallraff, and C. Eichler, Rapid high-fidelity multiplexed readout of superconducting qubits, \textit{Phys. Rev. Appl.} \textbf{10}, 034040 (2018).

\bibitem{broadbandFF}
N. T. Bronn, Y. Liu, J. B. Hertzberg, A. D. Córcoles, A. A. Houck, J. M. Gambetta, and J. M. Chow, Broadband filters for abatement of spontaneous emission in circuit quantum electrodynamics, \textit{Appl. Phys. Lett.} \textbf{107}, 172601 (2015).

\bibitem{Spring1}
P. A. Spring, L. Milanovic, Y. Sunada, S. Wang, A. F. van Loo, S. Tamate, and Y. Nakamura, Fast multiplexed superconducting qubit readout with intrinsic Purcell filtering, arXiv preprint, arXiv:2409.04967 (2024).

\bibitem{Rahamim2017Double-sided}
J. Rahamim, T. Behrle, M. J. Peterer, A. Patterson, P. A. Spring, T. Tsunoda, R. Manenti, G. Tancredi, and P. J. Leek, Double-sided coaxial circuit QED with out-of-plane wiring, \textit{Appl. Phys. Lett.} \textbf{110}, 222602 (2017).

\bibitem{spring}
P. A. Spring, S. Cao, T. Tsunoda, G. Campanaro, S. Fasciati, J. Wills, M. Bakr, V. Chidambaram, B. Shteynas, L. Carpenter, \textit{et al.}, High coherence and low cross-talk in a tileable 3D integrated superconducting circuit architecture, \textit{Science Advances} \textbf{8}, eabl6698 (2022).

\bibitem{her}
J. E. Johnson, C. Macklin, D. H. Slichter, R. Vijay, E. B. Weingarten, J. Clarke, and I. Siddiqi, Heralded state preparation in a superconducting qubit, \textit{Phys. Rev. Lett.} \textbf{109}, 050506 (2012).

\bibitem{dephasing}
J. M. Gambetta, A. Blais, D. I. Schuster, A. Wallraff, L. Frunzio, J. Majer, M. H. Devoret, S. M. Girvin, and R. J. Schoelkopf, Qubit-photon interactions in a cavity: Measurement-induced dephasing and number splitting, \textit{Phys. Rev. A} \textbf{74}, 042318 (2006).


\end{thebibliography}
\end{document}